\documentclass[]{article}
\usepackage[]{graphicx}

\begin{document}

\title{Propagation of localized optical waves in media with dispersion, in dispersionless media and in vacuum.
Non-diffracting  pulses}

\author{Lubomir M. Kovachev\\
Institute of Electronics, Bulgarian Academy of Sciences,\\
Tzarigradcko shossee 72,1784 Sofia, Bulgaria}
\date{}
\maketitle
\begin{abstract}

The applied method of the amplitude envelopes give us the
possibility to describe a new class of amplitude equations governing
the propagation of optical pulses in media with dispersion,
dispersionless media and vacuum.  We normalized these amplitude
equations and obtained five dimensionless parameters that determine
different linear and nonlinear regimes of the optical pulses
evolution. Surprisingly, in linear regime the normalized amplitude
equations for media with dispersion and the amplitude equations in
dispersionless media and vacuum are equal with precise - material
constants. The equations were solved for pulses with low intensity
(linear regime), using the method of Fourier transforms. One
unexpected new result was the relative stability of light bullets
and light disks and the significant reduction of their diffraction
enlargement. It is important to emphasize here  the case of light
disks, which turn out to be practically diffractionless over
distances of more than hundred kilometers.

\end{abstract}

\section{Introduction}

Laser physics experimentalist have reached the point of producing
without particular difficulties few picosecond or femtosecond (fsec)
optical pulses with equal duration in the $x$, $y$ and $z$
directions, the so called Light Bullets (LB), and fs optical pulses
with relatively large transverse and small longitudinal size (Light
Disks or LD). The evolution of so generated LB and LD in linear or
nonlinear regime is quite different from the propagation of light
filaments and light beams and has drawn the researchers' attention
with their unexpected dynamics behavior. For example, in \cite{Ruiz}
self-channeling of femtosecond pulses in air below the nonlinear
collapse threshold, i.e. in linear regime, was observed. This is in
contradiction with the well known paraxial approximation, which
gives good agreement between theory and experiment for optical beams
or long pulses with picosecond time duration. The main
characteristic of the paraxial approximation is the use of a
transverse Laplacian and the neglecting of the longitudinal second
derivative of the amplitude function in space. This leads to
Fresnel's type of diffraction for optical beams and long pulses in
linear regime. Various different nonlinear generalizations of the
paraxial approximation have been suggested
 based on the so-called spatio-temporal evolution equation
\cite{SIL,Kiv,Pon}. We suggest here another kind of nonparaxial
generalization of the amplitude equations in dependence of the
characteristics of the media ($\varepsilon(\omega)$ and $n_2 $), the
spectral region ($k_0$ and $ \omega_0$), and the shape of the pulse.
To provide a correct analysis, we need to introduce the different
spatial size of the optical pulses in the normalized equations.
The initial longitudinal spatial size is determined from the initial
time duration by the relations $z_0=v_gt_0=ct_0/n_g$, where $v_g$ is
the group velocity and $n_g$ is the group velocity refractive index.
For example, the typical transverse size of laser pulses is
evaluated from $r_{\perp}=1-5 $ $mm$ to $r_{\perp} =100$ $\mu m$,
while for the pulses' longitudinal size or the spatial period of
modulated waves, there are mainly three possibilities: a. Pulses
from nanoseconds to $30-40$ picoseconds. For such pulses
$z_0>>r_{\perp}$ and their shape is closer to light filaments
(LF). b. Pulses from few picoseconds to $300$-$400$ fs. In this case
$z_0\sim r_{\perp}$ and the pulses resemble LB. When the transverse
and longitudinal parts are approximately equal (case b), it is not possible to ignore
the second derivative along the $z$ direction in the amplitude equation.
c. Pulses from $200-300$ fs to $10-15$ fs and less.
In this case the transverse size exceeds the longitudinal one
$z_0<<r_{\perp}$ and the pulses' initial shape can be approximated with
light disks. The propagation of such a type of short optical pulses in
nonlinear media must be characterized not only by a different kind
of dimensions in the $x$, $y$ and $z$ directions, but also by a
non-stationary optical linear and nonlinear response of the media,
which lead to linear and nonlinear dispersion of different orders.
For example, the diffraction length $z_{diff}=k_0r_{\bot}^2$ of LB
and LD is of the order of the dispersion length $z_{dis}=t_0^2/k" $ for
solids materials and liquids in the UV region. This is why we should
expect that the equations governing the propagation of such types of
pulses, LB and LD, would be different from the well known paraxial
approximated equations, which describe the evolution of optical beams
\cite{CHIAO,TAL,KEL} and nanosecond or hundred picosecond pulses
\cite{ASX,ASC}. The inclusion of a longitudinal second derivative in
space and of second-order dispersion terms, which are not small for
LB and LD, changes the main result dramatically: even in the case of
a linear regime of propagation, the LB generated are relatively stable
and the LD are extremely stable in comparison with long pulses and light
beams.

\section{From the Maxwell's equations of a source-free, dispersive, nonlinear
 Kerr type medium to the amplitude equation}

 The propagation of ultra-short laser pulses in isotropic
 media, can be characterized with the following dependence of the polarization of first
 $\vec P_{lin}$ and third $\vec P_{nl}$ order on the electrical field $\vec E$:

\begin{eqnarray}
\label{eq6} \vec P_{lin} = \int\limits_{-\infty}^{t} {\left(\delta
(t-\tau) +4\pi\chi^{\left(1\right)} \left(t-\tau\right)\right)\vec
E\left(\tau, r\right)}d\tau =\nonumber\\
\int\limits_{-\infty}^{t} {\varepsilon \left(t-\tau\right)\vec
E\left(\tau, x, y, z\right)}d\tau,
\end{eqnarray}

\begin{eqnarray}
\label{pnl} \vec P_{nl}^{(3)} =3\pi
\int\limits_{-\infty}^{t}\int\limits_{-\infty}^{t-\tau_1}\int\limits_{-\infty}^{t-\tau_1-\tau_2}
\chi^{\left(3\right)}
\left(t-\tau_1,t-\tau_2,t-\tau_3\right)\nonumber\\
\times\left(\vec{E}(\tau_1,r)\cdot\vec E^{*}(\tau_2,r)\right)\vec
E(\tau_3,r) d\tau_1d\tau_2d\tau_3,
\end{eqnarray}
where $\chi^{(1)}$ and $\varepsilon$ are the linear electric
susceptibility and the dielectric constant, $\chi^{(3)}$ is the
nonlinear susceptibility of third order, and we denote $r=(x,y,z)$.
We use this expression of the nonlinear polarization (\ref{pnl}), as
we will investigate only linearly or only
circularly polarized light. The Maxwell's equations in this case
become:

\begin{eqnarray}
\label{eq1}
\nabla \times \vec E =
- \frac{1}{c}\frac{\partial \vec B}{\partial t},
\end{eqnarray}

\begin{eqnarray}
\label{eq2}
\nabla \times \vec H =
  \frac{1}{c}\frac{\partial \vec D}{\partial t},
\end{eqnarray}

\begin{equation}
\label{eq3}
\nabla \cdot \vec D = 0,
\end{equation}

\begin{equation}
\label{eq4}
\nabla \cdot \vec B = \nabla \cdot \vec H = 0,
\end{equation}

\begin{equation}
\label{eq5} \vec B = \vec H,\ \vec D = \vec P_{lin}  +  \vec P_{nl},
\end{equation}
where $\vec E$ and $\vec H$ are the electric and magnetic fields intensity
, $\vec D$ and $\vec B$ are the electric and magnetic
inductions. We should point out here that these equations
are also valid when the time duration of the optical pulses $t_0$
is greater than the characteristic response time of the media
$\tau_0$ ($t_0>>\tau_0$), and when the time duration of the pulses is
of the order of time response of the media ($t_0\leq\tau_0$).

Taking the curl of equation (\ref{eq1}) and using (\ref{eq2}) and
(\ref{eq5}), we obtain:

\begin{eqnarray}
\label {eq8} \nabla \left(\nabla\cdot\vec E\right) -\Delta\vec E=
-\frac{1}{c^2}\frac{\partial^2\vec D}{\partial t^2},
\end{eqnarray}
where $\Delta \equiv \nabla ^2$ is the Laplace operator.

Equation (\ref{eq8}) is derived without using the third
Maxwell's equation. Using equation (\ref{eq3}) and the expression
for the linear and nonlinear polarizations (\ref{eq6}) and
(\ref{pnl}), we can estimate the second term in equation
(\ref{eq8}) for arbitrary localized vector function of the
electrical field. It is not difficult  to show that for localized
functions in nonlinear media with and without dispersion  $\nabla
\cdot\vec {E}\cong 0$ and equation (\ref{eq8}) we can write:
\begin{eqnarray}
\label {lap} \Delta\vec E= \frac{1}{c^2}\frac{\partial^2\vec
D}{\partial t^2},
\end{eqnarray}
We will now express the electrical field in linear and nonlinear
polarization on the right-hand side of (\ref{lap}) with it's Fourier
integral:

\begin{eqnarray}
\label {F1} \vec E\left(r,t\right)= \int\limits_{-\infty}^{+\infty}
\hat{\vec E}\left(r,\omega\right) \exp{\left(-i\omega
t\right)}d\omega,
\end{eqnarray}

We thus obtain:

\begin{eqnarray}
\label{lapint} \Delta\vec E= \frac{1}{c^2}\frac{\partial^2}{\partial
t^2} \left(\int\limits_{-\infty}^{t} \varepsilon(t-\tau) \hat{\vec
E}\left(r,\omega\right)\exp{(-i\omega\tau)}d\omega d\tau\right)
\nonumber\\
+\frac{3\pi}{c^2}\frac{\partial^2}{\partial t^2}
\int\limits_{-\infty}^{t}\int\limits_{-\infty}^{t-\tau_1}
\int\limits_{-\infty}^{t-\tau_1-\tau_2}\chi^{\left(3\right)}
\left(t-\tau_1,t-\tau_2,t-\tau_3\right)\left|\hat{\vec
E}\left(r,\omega\right)\right|^2 \hat{\vec
E}\left(r,\omega\right)\nonumber\\
\times\exp{\left(-i\left(\omega(\tau_1-\tau_2+\tau_3)\right)\right)}
d\tau_1d\tau_2d\tau_3.
\end{eqnarray}
The causality principle imposes the following conditions on the
response functions:

\begin{eqnarray}
\label{caus} \varepsilon(t-\tau)=0;\ \chi^{\left(3\right)}
\left(t-\tau_1,t-\tau_2,t-\tau_3\right)=0, \nonumber\\
 t-\tau<0;\ t-\tau_i<0;\ i=1,2,3.
\end {eqnarray}

This is why we can extend the upper integral boundary to infinity
and use the standard Fourier transform \cite{MN}:

\begin{eqnarray}
\label{intcs} \int\limits_{-\infty}^{t}
{\varepsilon(\tau-t)\exp{(i\omega\tau)}d\tau}=\int\limits_{-\infty}^{+\infty}
{\varepsilon(\tau-t)\exp{(i\omega\tau)}d\tau}\\
\int\limits_{-\infty}^{t}\int\limits_{-\infty}^{t-\tau_1}\int\limits_{-\infty}^{t-\tau_1-\tau_2}
\chi^{\left(3\right)} \left(t-\tau_1,t-\tau_2,t-\tau_3\right)
d\tau_1d\tau_2d\tau_3=\nonumber\\
\int\limits_{-\infty}^{+\infty}\int\limits_{-\infty}^{+\infty}\int\limits_{-\infty}^{+\infty}
\chi^{\left(3\right)} \left(t-\tau_1,t-\tau_2,t-\tau_3\right)
d\tau_1d\tau_2d\tau_3.
\end {eqnarray}

The spectral representation of the linear optical susceptibility
$\hat{\varepsilon_{0}}(\omega)$ is connected to the non-stationary
optical response function by the following Fourier transform:

\begin{eqnarray}
\label{chi1} \hat{\varepsilon}(\omega)\exp{(-i\omega t)}
=-\int\limits_{-\infty}^{+\infty}
{\varepsilon(t-\tau)\exp{(-i\omega\tau)}d\tau}.
\end{eqnarray}
The expression for the spectral representation of the non-stationary
nonlinear optical susceptibility $\hat{\chi}^{(3)}$ is similar :

\begin{eqnarray}
\label{chi3} \hat{\chi}^{(3)}(\omega)\exp{(-i\omega t)}=-
\int\limits_{-\infty}^{+\infty}\int\limits_{-\infty}^{+\infty}
\int\limits_{-\infty}^{+\infty}\chi^{\left(3\right)}
\left(t-\tau_1,t- \tau_2,t-\tau_3\right)\nonumber\\
\times\exp{\left(-i\left(\omega(\tau_1-\tau_2+\tau_3)\right)\right)}
d\tau_1d\tau_2d\tau_3.
\end{eqnarray}
Thus, after brief calculations, equation (\ref{lapint}) can be
represented as

\begin{eqnarray}
\label{lapomeg} \Delta\vec E=-\int\limits_{-\infty}^{\infty}
\frac{\omega^2\hat{\varepsilon}(\omega)}{c^2} \hat{\vec
E}\left(r,\omega\right)\exp{(-i\omega t)}d\omega
\nonumber\\
+
\int\limits_{-\infty}^{\infty}\frac{\omega^2\hat{\chi}^{\left(3\right)}
\left(\omega\right)}{c^2}\left|\hat{\vec
E}\left(r,\omega\right)\right|^2 \hat{\vec
E}\left(r,\omega\right)\exp{\left(-i\left(\omega t\right)\right)}
d\omega.
\end{eqnarray}
We now define the square of the linear $k^2$ and the nonlinear $k_{nl}^2$ wave
vectors as well as the nonlinear refractive index $n_2$ with the
expressions:
\begin{eqnarray}
k^2=\frac{\omega^2\hat{\varepsilon}\left(\omega\right)}{c^2},
\end{eqnarray}

\begin{eqnarray}
k_{nl}=\frac{3\pi\omega^2\hat{\chi}^{(3)}\left(\omega\right)}{c^2}=k^2n_2,
\end{eqnarray}
where
\begin{eqnarray}
n_2(\omega)=\frac{3\pi\hat{\chi}^{(3)}\left(\omega\right)}{\hat{\varepsilon}\left(\omega\right)}.
\end{eqnarray}
In terms of these quantities, the equation (\ref{lapomeg}) can be
expressed by:

\begin{eqnarray}
\label{lapk} \Delta\vec E=-\int\limits_{-\infty}^{\infty}
k^2(\omega) \hat{\vec E}\left(r,\omega\right)\exp{(-i\omega
t)}d\omega
\nonumber\\
-
\int\limits_{-\infty}^{\infty}k^2(\omega)n_2(\omega)\left|\hat{\vec
E}\left(r,\omega\right)\right|^2 \hat{\vec
E}\left(r,\omega\right)\exp{\left(-i\left(\omega t\right)\right)}
d\omega.
\end{eqnarray}
Let us introduce here the amplitude function $\vec A(r,t)$ for the
electrical field $\vec E(r,t)$:

\begin{eqnarray}
\label{a1}
 \vec E\left(r,t\right)=\vec {A}\left(r,t\right)\exp{\left(i(k_0z-\omega_0t)\right)}+c.c,
\end{eqnarray}
the Fourier transform of the amplitude function $\hat{\vec
A}(r,\omega-\omega_0)$:

\begin{eqnarray}
\label {FA} \vec A\left(r,t\right)= \int\limits_{-\infty}^{+\infty}
\hat{\vec A}\left(r,\omega-\omega_0\right)
\exp{\left(-i(\omega-\omega_0) t\right)}d\omega,
\end{eqnarray}
and the following relation between the Fourier transform of
electrical field and the Fourier transform of the amplitude
function:

\begin{eqnarray}
\label{hata}
 \hat{\vec E}\left(r,\omega\right)\exp(-i\omega t)=\exp\left(-i\left(k_0z-\omega_0 t\right)\right)
 \hat{\vec
 {A}}\left(r,\omega-\omega_0\right)\exp{\left(i(\omega-\omega_0)t\right)},
\end{eqnarray}
where $\omega_0$ and $k_0 $ are the optical frequency and the wave
number of the optical field, respectively. Since we investigate
optical pulses, we assume that the amplitude function and its
Fourier expression are time - and frequency-localized. Substituting
(\ref{a1}),(\ref{FA}) and (\ref{hata}) into equation (\ref{lapk}) we
finally obtain the following nonlinear integro-differential
amplitude equation:

\begin{eqnarray}
\label{ampk} \Delta\vec A(r,t) + 2ik_0\frac{\partial\vec
A(r,t)}{\partial
z}- k_0^2\vec A(r,t)=\nonumber \\
-\int\limits_{-\infty}^{\infty}
k^2(\omega)\left(1+n_2(\omega)\left|\hat{\vec
A}\left(r,\omega-\omega_0\right)\right|^2 \right) \hat{\vec
A}\left(r,\omega-\omega_0\right)\exp{(-i(\omega-\omega_0) t)}d\omega
\end{eqnarray}

The equation (\ref{ampk}) was derived with only one restriction,
namely, that the amplitude function and its Fourier expression be
localized functions. This is why, if we know the analytical
expression of $k^2(\omega)$ and $n_2(\omega)$, the Fourier integral
on the right- hand side of (\ref{ampk}) is a finite integral away
from resonances. In this way we can also investigate optical pulses
with time duration $t_0$ of the order of the optical period
$T_0=2\pi/\omega_0$. Generally speaking, using the nonlinear
integro-differential amplitude equation (\ref{ampk}) we can also
investigate wave packets with time duration of the order of the
optical period, as well as wave packets with a large number of
harmonics under the pulse. The nonlinear integro-differential
amplitude equation (\ref{ampk}) can be written as a nonlinear
differential equation for the Fourier retransform of the amplitude
function $\hat{\vec A}$, after we apply the time Fourier
transformation (\ref{FA}) to the left-hand side of (\ref{ampk}) :

\begin{eqnarray}
\label{ampf} \Delta\hat{\vec A}\left(r,\omega-\omega_0\right)+
2ik_0\frac{\partial\hat{\vec
A}\left(r,\omega-\omega_0\right)}{\partial z}\nonumber \\
+\left(\left(1+n_2(\omega)\left|\hat{\vec
A}\left(r,\omega-\omega_0\right)\right|^2
\right)k^2(\omega)-k_0^2(\omega_0)\right) \hat{\vec
A}\left(r,\omega-\omega_0\right)=0.
\end{eqnarray}

We note here the well-known fact that the Fourier component of the
amplitude function in equation (\ref{ampf}) depends on the spectral
difference $\triangle\omega=\omega-\omega_0$, rather than on the
frequency, as is the case for the electrical field.

\section{From amplitude equation to the slowly varying equation of amplitudes (SVEA) }

The equation (\ref{ampk}) is obtained without imposing any restrictions on the
square of the linear $k^2(\omega)$ and nonlinear
$k^2_{nl}=k^2(\omega)n_2(\omega)$ wave vectors. To obtain SVEA, we will
restrict our investigation to the cases when it is possible to
approximate $k^2$ and $k_{nl}^2$ as a power series with respect to the frequency
difference $\omega-\omega_0$ as:

\begin{eqnarray}
\label{a9}
 k^2\left(\omega\right)=
\frac{\omega ^2\hat{\varepsilon_0}\left(\omega\right)}{c^2} =
 k^{2}\left(\omega_0\right) +
\frac{\partial\left(k^2\left(\omega_0\right)\right)}{\partial\omega_0}
\left(\omega-\omega_0\right) \nonumber\\
+ \frac{1}{2}
\frac{\partial^2\left(k^2\left(\omega_0\right)\right)}{\partial
\omega_0^2} \left(\omega-\omega_0\right)^2 + ...,
\end{eqnarray}
\begin{eqnarray}
\label{nla9}
 k_{nl}^2\left(\omega\right)=
\frac{\omega ^2\hat{\chi}^{(3)}\left(\omega\right)}{c^2} =
k_{nl}^{2} \left(\omega_0\right) +
\frac{\partial\left(k_{nl}^2\left(\omega_0\right)\right)}{\partial\omega_0}
\left(\omega-\omega_0\right)+...
\end{eqnarray}
To obtain SVEA in second approximation to the linear dispersion and in
first approximation to the nonlinear dispersion, we must cut off these
series to the second derivative term for the linear wave vector and to
the first derivative term for the nonlinear wave vector. This is
possible only if the series (\ref{a9}) and (\ref{nla9}) are strongly
convergent. Than, the main value in the Fourier integrals in
equation (\ref{ampk}) yields the first and second derivative terms in
(\ref{a9}), and the zero and first derivative terms in (\ref{nla9}).
The first term in (\ref{a9}) cancels the last term on the left-hand side of the
equation (\ref{ampk}). The convergence of the series
(\ref{a9}) and (\ref{nla9}) for spectrally limited pulses propagating
in the transparency UV and optical regions of solids materials,
liquids and gases, depends mainly on the number of harmonics under
the pulses \cite {KAM}. For wave packets with more than 10 harmonics
under the envelope, the series (\ref{a9}) is strongly convergent, and the
third derivative term (third order of dispersion) is smaller than
the second derivative term (second order of dispersion) by three to
four orders of magnitude. In this case we can cut off the series to the second
derivative term in (\ref{a9}), as the higher terms in the series
contribute very little to the Fourier integral in equation
(\ref{ampk}). In the case of 3-7 harmonics under the pulse, the
series (\ref{a9}) is weakly convergent and we must take into account the dispersion terms of
higher orders of as small parameters . In the case of wave
packets with only one or two harmonics under the pulse the
series (\ref{a9}) is divergent. This is the reason why the SVEA does not
govern the dynamics of wave packets with time duration of the order
of the optical period. As we pointed out above, this restriction is
not valid for the amplitude equation (\ref{ampk}) and we can use it
to investigate such kind of pulses. Substituting the series (\ref{a9}) and
(\ref{nla9}) in (\ref{ampk}) and bearing in mind the expressions for the
time-derivative of the amplitude function, the SVEA of second order with respect to
the linear dispersion and first order with respect to the nonlinear dispersion is
expressed in the following  in next form:

\begin{eqnarray}
\label{sve} \Delta\vec A + 2ik_0\frac{\partial\vec A}{\partial z}+
2ik_0k'\frac{\partial\vec A}{\partial
t}=\nonumber\\
\left(k_0k"+k'^2\right)\frac{\partial^2\vec
A}{\partial t^2}-k_{0nl}^2\left|\vec A\right|^2 \vec
A-2ik_{0nl}k'_{0nl}\frac{\partial\left|\vec A\right|^2 \vec
A}{\partial t}
\end{eqnarray}
We will now define other important constants connected with the wave
packets carrier frequency: linear wave vector $k_0\equiv
k(\omega_0)=\omega_0\sqrt{\varepsilon(\omega_0)}/c$ linear
refractive index $n(\omega_0)=\sqrt{\varepsilon(\omega_0)}$,
nonlinear refractive index
$n_2(\omega_0)=3\pi\chi^{(3)}(\omega_0)/\varepsilon(\omega_0)$,
group velocity:

\begin{eqnarray}
\label{vgr}
v(\omega_0)=\frac{1}{k'}=\frac{c}{\sqrt{\varepsilon(\omega_0)}+
\frac{\omega_0}{2}\sqrt{\frac{1}{\varepsilon}}\frac{\partial\varepsilon}{\partial\omega}},
\end{eqnarray}
nonlinear addition to the group velocity $k_{0nl}^2$:

\begin{eqnarray}
\label{vgrn} \left(k^2_{0nl}\right)=\frac{2k_0n_2}{v}+k_0^2
\frac{\partial n_2}{\partial\omega},
\end{eqnarray}
and dispersion of the group velocity $k"(\omega_0)=\partial^2
k/\partial\omega^2_{\omega=\omega_0}.$ All these quantities allow
a direct physical interpretation and we will, therefore, rewrite equation
(\ref{sve}) in a form consistent with these constants:

\begin{eqnarray}
\label{svea}  -i\left[\frac{\partial\vec A}{\partial t}+
v\frac{\partial\vec A}{\partial
z}+\left(n_2+\frac{k_0v}{2}\frac{\partial n_2}{\partial
\omega}\right)\frac{\partial\left(\left|\vec A\right|^2 \vec
A\right)}{\partial t}\right]
=\nonumber\\
\frac{v}{2k_0}\Delta\vec A
-\frac{v}{2}\left(k"+\frac{1}{k_0v^2}\right)\frac{\partial^2\vec
A}{\partial t^2}+\frac{k_{0}vn_2}{2}\left|\vec A\right|^2 \vec A
\end{eqnarray}
This equation can be considered to be SVEA of second approximation
with respect to the linear dispersion and of first approximation, to the nonlinear
dispersion (nonlinear addition to the group velocity). It includes
the effects of translation in z direction with group velocity $v$,
self-steepening, diffraction, dispersion of second order and
self-action terms.

\section{Propagation of ultrashort optical pulses in vacuum and dispersionless media}
The theory of light envelopes is not restricted only to the cases of
non-stationary optical (and magnetic) response. Even in vacuum,
where $\varepsilon=1$  and $\vec P_{nl}=0$ we can write an amplitude
equation by applying solutions of the kind (\ref{a1}) to the wave equation (\ref{eq8}).
We thus obtain the following linear equation for the
amplitude envelope of the electrical field:

\begin{eqnarray}
\label{vac}  -i\left(\frac{\partial\vec A}{\partial t}+
c\frac{\partial\vec A}{\partial z}\right) = \frac{c}{2k_0}\Delta\vec
A -\frac{1}{2k_0c}\frac{\partial^2\vec A}{\partial t^2}.
\end{eqnarray}
The Vacuum Linear Amplitude Equation (VLAE) (\ref{vac}) is obtained
directly from the wave equation without any restrictions by using
the series of the square of the wave vector, as is the case of a
media with dispersion, where it is required that the series
(\ref{a9}) be strongly convergent. This is why equation (\ref{vac})
describes both amplitudes with many harmonics under the pulse, and
amplitudes with only one or a few harmonics under the envelope. It
is obvious that the envelope $\vec A$ in equation (\ref{vac}) will
propagate with the speed of light in vacuum $c$. The equation
(\ref{vac}) is valid also for transparent media with stationary
optical response $\varepsilon=const$. In this case, the propagating
constant will be $v=c/\sqrt{\varepsilon\mu}$.

\section {SVEA and VLAE in a normalized form}

Starting from Maxwell's equations for media with non-stationary
linear and nonlinear response, we obtained an amplitude equation
and a SVEA using only two restrictions, which are physically
acceptable for ultra-short pulses. Having adopted the first restriction, namely,
investigation of localized in time and space amplitude
functions only, we introduced the amplitude equation (\ref{ampk}).
Following the second restriction, i.e., limiting ourselves with the case of a large number of
harmonics under the localized envelopes, we obtained the SVEA
(\ref{svea}). As it was pointed out in the previous section, these
restrictions do not affect the VLAE (\ref{vac}). The next step is
writing SVEA (\ref{svea}) and VLAE (\ref{vac}) in dimensionless variables
and estimating the influence of the different differential terms. In
this case, the coefficients in front of the differential operators in
(\ref{svea}) and (\ref{vac}) will be numbers of different orders,
depending on the media $n$ and $n_2$, the spectral region of
propagation $k_0$ and $\omega_0$, the field intensity
$\left|A_0\right|^2$, and the initial shape of the pulses, namely, optical
filament $r_\bot<<z_0$, optical bullets $r_\bot\approx z_0$ or
optical disks $r_\bot>>z_0$. With $r_\bot$ we denote here the
initial transverse dimension, "the spot" of the pulse, and with
$z_0$ we denote the initial longitudinal dimension, which is simply
the spatial analog of the initial time duration $t_0$, determined by
the relation $z_0=vt_0$ or $z_0=ct_0$ in the vacuum case. The SVEA
(\ref{svea}) and VLAE (\ref{vac}) are written in a Cartesian
laboratory coordinate system. To investigate the dynamics of optical
pulses on long distances, it is convenient to rewrite these equations
in a Galilean coordinate system, where the new reference frame moves
with the group velocity, $t' = t; z' = z - vt$ for equation
(\ref{svea}):

\begin{eqnarray}
\label{gal} -i\left(\frac{\partial\vec A}{\partial t'}+
\left(n_2+\frac { k_0v}{2}\frac{\partial
n_2}{\partial\omega}\right)\left(\frac{\partial \left(\left|\vec
A\right|^2\vec A\right)}{\partial t'}-\frac{\partial
\left(\left|\vec A\right|^2\vec A\right)}{\partial z'}
\right)\right)=
 \frac{v}{2k_0}\Delta_{\bot}\vec A- \nonumber\\
\frac{v^3k_0^{"}}{2}\frac{\partial^2\vec A}{\partial z'^2}
-\frac{v}{2}\left(k"+\frac{1}{k_0v^2}\right)\left(\frac{\partial^2\vec
A}{\partial t'^2}-2v\frac{\partial^2\vec A}{\partial t'\partial
z'}\right)+ \frac{n_2 k_0 v}{2}\left|\vec A\right|^2\vec A,
\end{eqnarray}
and with velocity of light for equation (\ref{vac}), $t' = t; z' = z
- ct$:

\begin{eqnarray}
\label{galvac} -i\frac{\partial\vec A}{\partial t'}=
\frac{c}{2k_0}\Delta_{\bot}\vec A-
\frac{1}{2k_0c}\frac{\partial^2\vec A}{\partial
t'^2}+\frac{1}{k_0}\frac{\partial^2\vec A}{\partial t'\partial z'},
\end{eqnarray}
and with $\Delta_{\bot}=\frac{\partial^2}{\partial x^2} +
\frac{\partial ^2}{\partial y^2}$ we denote the transverse
Laplacian. We define the following dimensionless variables, connected
with the initial amplitude and with the spatial and temporal dimensions of the
pulses with the relations:

\begin{eqnarray}
\label{eq12} \vec A=A_0\vec A";\ x=r_{\bot} x";\ y=r_{\bot}y";\
z'=z_0z";\ t'=t_0t";\ z=z_0z";\ t=t_0t".
\end{eqnarray}
We substitute these variables in (\ref{svea}), (\ref{gal}), (\ref{vac})
and (\ref{galvac}) and, making use of the expressions for the diffraction
$ z_{dif}=k_0r_\bot^2$ and dispersion $z_{disp}=t_0^2/k"$ lengths,
we obtain the next five dimensionless parameters in front of the
differential terms in the equations (\ref{svea}), (\ref{gal})
(\ref{vac}) and (\ref{galvac}):

\begin{eqnarray}
\label{norm}
 \alpha=k_0z_0;\ \delta^2=\frac{r_{\bot}^2}{z_0^2};\ \beta=\frac {z_{dif}}{z_{disp}};\
\gamma=k_0^2r_0^2n_2\left|A_0\right|^2;\nonumber\\
\gamma_1=2k_0r_0n_2\left|A_0\right|^2\left(n_2+\frac {
k_0v}{2}\frac{\partial n_2}{\partial\omega}\right).
\end{eqnarray}
In the new dimensionless variables and constants, the equations
(\ref{svea}), (\ref{gal}), (\ref{vac}) and (\ref{galvac}) can be
represented as (the seconds of dimensionless variables are omitted for simplicity):

Case a. SVEA (\ref{svea}) in a laboratory frame ("Laboratory")
\begin{eqnarray}
\label{E13}
 -2i\alpha\delta^2\left(\frac{\partial\vec A}{\partial t}+
\frac{\partial\vec A}{\partial z}+\gamma_1\frac{\partial
\left(\left|\vec A\right|^2\vec A\right)}{\partial t}\right)=
\Delta_{\bot}\vec A +\delta^2\left(\frac{\partial^2\vec A}{\partial
z^2}-\frac{\partial^2\vec A}{\partial t^2}\right)
-\nonumber\\
\\
 \beta\frac{\partial^2\vec A}{\partial
t^2} + \gamma \left|\vec A\right|^2\vec A. \nonumber
\end{eqnarray}

Case b. SVEA (\ref{gal}) in a frame moving with the group velocity:

\begin{eqnarray}
\label{G13}
 -2i\alpha\delta^2\left(\frac{\partial\vec A}{\partial t'}+
\gamma_1\left(\frac{\partial \left(\left|\vec A\right|^2\vec
A\right)}{\partial t'}-\frac{\partial \left(\left|\vec
A\right|^2\vec A\right)}{\partial z'} \right)\right)=
\Delta_{\bot}\vec A -\beta\frac{\partial^2\vec A}{\partial z'^2}
-\nonumber\\
\\
 \left(\beta +\delta^2\right)\left(\frac{\partial^2\vec A}{\partial
t'^2}-2\frac{\partial^2\vec A}{\partial t'\partial z'}\right) +
\gamma \left|\vec A\right|^2\vec A, \nonumber
\end{eqnarray}

Case c. VLAE (\ref{vac}) in a laboratory frame:
\begin{eqnarray}
\label{V13}  -2i\alpha\delta^2\left(\frac{\partial\vec A}{\partial
t}+ \frac{\partial\vec A}{\partial z}\right) = \Delta_{\bot}\vec A
+\delta^2\left(\frac{\partial^2\vec A}{\partial
z^2}-\frac{\partial^2\vec A}{\partial t^2}\right).
\end{eqnarray}

Case d. VLAE (\ref{galvac}) in a Galilean frame:
\begin{eqnarray}
\label{VG13} -2i\alpha\delta^2\frac{\partial\vec A}{\partial t'}=
\Delta_{\bot}\vec A- \delta^2\left(\frac{\partial^2\vec A}{\partial
t'^2}-\frac{\partial^2\vec A}{\partial t'\partial z'}\right).
\end{eqnarray}
We obtain equal dimensionless constants in front of the differential
terms in both the "Laboratory" and "Galilean" frames. This gives us the
possibility to investigate and estimate simultaneously the different
terms in the normalized equations (\ref{E13}), (\ref{G13}), (\ref{V13})
and(\ref{VG13}). We will now discuss these constants in more detail,
as they play a significant role in determining the different pulse
propagation regimes.

- The first constant $\alpha=k_0z_0=2\pi z_0/\lambda_0$ determines
with precision $2\pi$ the "number of harmonics" on a FWHM level of the pulses.
Since we use the slowly varying amplitude approximation, $\alpha$ is always a large number ($\alpha>>1$).

- The second constant $\delta^2=r_\bot^2/z_0^2$ determines the
relation between the initial transverse and longitudinal size of the
optical pulses. This parameter distinguishes the case of light
filaments (LF) $\delta^2=r_\bot^2/z_0^2<<1$ from the case of LB
$\delta^2=r_\bot^2/z_0^2\cong1$ and the case of light disks LD
$\delta^2=r_\bot^2/z_0^2>>1$. For light filaments $\delta^2<<1$ and
we can neglect the differential terms with coefficient $\delta^2$.
It is not  difficult to see that in this case the SVEA (\ref{E13}),
(\ref{G13}) and VLAE (\ref{G13}), (\ref{V13}) can be transformed to
the standard paraxial approximation of the linear and nonlinear
optics. If we set the possible values of the optical pulses'
transverse dimensions at $3-4 mm > r_{\bot} >100 \mu m$, we can
directly obtain above distinction in dependence of time duration of
the pulses. For light pulses with time duration $ns >t_0
> 40-50 ps$ we obtain $\delta^2=r_\bot^2/z_0^2<<1$ and we are in
regime of LF and paraxial approximation. In case of light pulses
with duration $t_0\approx 3-4 ps$ up to $500-600 fs$ it is possible
to reach $\delta^2=r_\bot^2/z_0^2\cong1$ and we are in regime of LB.
For pulses in the time range $300 fs-30 fs$ we can prepare the initial
shape of the pulses to satisfy the relation
$\delta^2=r_\bot^2/z_0^2>>1$ and thus reach the LD regime. It is
important to note here that the wave packets in visible and UV
ranges with time duration $t_0\geq 30 fs$ admit more than 10-15
optical harmonics under the pulse, so that we are still in
SVEA approximation. In the last two cases (LB and LD) the
differential terms with $\delta^2$ can't be ignored and the
equations (\ref{E13}), (\ref{G13}),(\ref{G13}) and(\ref{V13})
governed the propagation of pulses with initial form of LB and LD
are quite different than paraxial approximation.

-The third parameter is $\beta=k_0r_\bot^2/z_{dis}$, where
$z_{dis}=t_0/k"$, determines the relation between diffraction and
dispersion length. The dispersion parameter $k"$ in the visible and
UV transparency region of dielectrics has values from $k"\sim
10^{-31}$ $s^2/cm$  for gases and metal vapors up to $k"\sim
10^{-26}$ $s^2/cm$ for solid materials. It is convenient to express
this parameter using the product of the second constant $\delta^2$
and the parameter $\beta_1=k_0v^2k"$ by the relation
$\beta=\beta_1\delta^2$. For the typical values of the dispersion
$k"$ in the visible and UV region listed above, the dimensionless
parameter $\beta_1$ is very small ($\beta_1<<1$) in the visible part
of the spectrum, while for optical pulses propagating in the UV
region in solids and liquids it may reach $\beta_1\propto 1$. The
parameter $\beta_1$ can be also negative and may reach
$\beta_1\simeq -1$ near electronic resonances and also near the
Langmuir frequency in electronic plasmas \cite{LMK}. It is no
difficult to see that only in the case $\beta_1\simeq -1$ and
$\delta^2\propto 1$ can we obtain the 3D+1 nonlinear Schrödinger
equation from SVEA (\ref{E13}) and (\ref{G13})

- The fourth and fifth constants
$\gamma=k_0^2r_{\perp}^2n_2\left|A_0\right|^2/2$ and
$\alpha\delta^2\gamma_1=\alpha\delta^2n_2\left|A_0\right|^2/2$ are
correspondingly the nonlinear coefficient and the coefficient of
nonlinear addition to the group velocity (coefficient in the front of
the first order nonlinear dispersion). It is easy to estimate
that for $\alpha>>1$ and $\delta^2\geq1$, we always have
$\gamma>>\alpha\delta^2\gamma_1$. In this paper we investigate
optical pulses with power near the critical threshold for self-focusing
$\gamma\cong1$ and less (linear regime) $\gamma<<1$. For these cases
the nonlinear addition to the group velocity is very small
($\alpha\delta^2\gamma_1<<1$) and from here to the end of this paper
we will neglect the terms with the first addition to the nonlinear
dispersion. The analysis of the dimensionless constants performed above leads us
to the following conclusion: The investigation of wave packets with power
near that for self-focusing $\gamma\propto1$ in the visible and UV region in
a media with dispersion are governed by the following SVEA equations:

Case a. Laboratory frame ("Laboratory")
\begin{eqnarray}
\label{LAB13}
 -2i\alpha\delta^2\left(\frac{\partial\vec A}{\partial t}+
\frac{\partial\vec A}{\partial z}\right)= \Delta_{\bot}\vec A
+\delta^2\frac{\partial^2\vec A}{\partial z^2}
 -\delta^2\left(\beta_1+1\right)\frac{\partial^2\vec A}{\partial
t^2} + \gamma \left|\vec A\right|^2\vec A.
\end{eqnarray}

Case b. Frame moving with group velocity ("Galilean"):

\begin{eqnarray}
\label{GAL13}
 -2i\alpha\delta^2\frac{\partial\vec A}{\partial t'}=
\Delta_{\bot}\vec A -\beta_1\delta^2\frac{\partial^2\vec A}{\partial
z'^2} - \delta^2\left(\beta_1 +1\right)\left(\frac{\partial^2\vec
A}{\partial t'^2}-2\frac{\partial^2\vec A}{\partial t'\partial
z'}\right) + \gamma \left|\vec A\right|^2\vec A.
\end{eqnarray}
Equations (\ref{LAB13}) and (\ref{GAL13}) differ from the paraxial
spatio-temporal evolution equations investigated in
\cite{SIL,Kiv,Pon,Chris} by the inclusion of a second derivative
along the $z$ direction, mixed term and additional second derivative
in time term. This leads to a quite different dynamics of the
ultrashort pulses; we will first investigate their propagation in
linear regime, when $\gamma<<1$.

\section{Analytical solutions of the linear SVEA and VLAE}
The behavior of long pulses is similar to that of optical beams, since
their propagation is governed by a similar kind of paraxial equations, as
pointed above. That is why we cannot expect the diffraction enlargement
of long pulses to be different from that of optical beams. The situation
with the LB and LD is different. Their propagation is governed by
the new spatio-temporal evolution equations in media with
non-stationary optical response - SVEA (\ref{LAB13}) and
(\ref{GAL13}), and by VLAE (\ref{vac}) and (\ref{galvac} in media with
linear stationary optical response (or vacuum). In this section we will
attempt to solve the equations (\ref{LAB13}), (\ref{GAL13}) in linear
regime ($\gamma<<1$) and will compare the solutions with the
solutions of the linear VLAE (\ref{vac}) and (\ref{galvac}).
Neglecting the small nonlinear terms in (\ref{LAB13}), (\ref{GAL13})
we arrive at:

a. Linear SVEA in a laboratory coordinate frame:

\begin{eqnarray}
\label{eqlin13}
 -2i\alpha\delta^2\left(\frac{\partial\vec A}{\partial t}+
\frac{\partial\vec A}{\partial z}\right)= \Delta_{\bot}\vec A
+\delta^2\frac{\partial^2\vec A}{\partial
z^2}-\delta^2\left(\beta_1+1\right)\frac{\partial^2\vec A}{\partial
t^2}.
\end{eqnarray}

b. Linear SVEA in a Galilean coordinate frame:

\begin{eqnarray}
\label{lin13}
 -2i\alpha\delta^2\frac{\partial\vec A}{\partial t'}=
\Delta_{\bot}\vec A
-\delta^2\left(\beta_1+1\right)\left(\frac{\partial^2\vec
A}{\partial t'^2}-2\frac{\partial^2\vec A}{\partial t'\partial
z'}\right)-\delta^2\beta_1\frac{\partial^2\vec A}{\partial z'^2}.
\end{eqnarray}
For comparison we will rewrite here the corresponding linear VLAE:

c. Linear VLAE  in a laboratory frame:

\begin{eqnarray}
\label{VV13}  -2i\alpha\delta^2\left(\frac{\partial\vec A}{\partial
t}+ \frac{\partial\vec A}{\partial z}\right) = \Delta_{\bot}\vec A
+\delta^2\left(\frac{\partial^2\vec A}{\partial
z^2}-\frac{\partial^2\vec A}{\partial t^2}\right).
\end{eqnarray}
d. Linear VLAE in a Galilean frame:

\begin{eqnarray}
\label{VVG13} -2i\alpha\delta^2\frac{\partial\vec A}{\partial t'}=
\Delta_{\bot}\vec A- \delta^2\left(\frac{\partial^2\vec A}{\partial
t'^2}-\frac{\partial^2\vec A}{\partial t'\partial z'}\right).
\end{eqnarray}
The important result here is that in linear regime the normalized
equations for media with dispersion (\ref{eqlin13}), (\ref{lin13})
identical equal with accuracy of one dimensionless dispersion parameter
$\beta_1$ with the equations for media without dispersion and vacuum
(\ref{VV13}), (\ref{VVG13}). In the equations (\ref{eqlin13}),
(\ref{lin13}) and (\ref{VV13}), (\ref{VVG13}) there are only three
dimensionless parameters, namely, $\alpha=k_0r_0>>1$, which determines the
"number of harmonics" on FWHM level, $\delta^2=r_0^2/z_0^2$, which determines
the relation between the longitudinal $z_0$ and the transverse $r_{\bot}$ size of the
optical pulse, and the dispersion parameter $\beta_1=k_0v^2k"$. As it
can be expected, the equations for ultra-short optical pulses in
vacuum and dispersionless media (\ref{VV13}), (\ref{VVG13}) become
identical with the equations with dispersion (\ref{eqlin13}),
(\ref{lin13}) when the dispersion parameter $ \beta_1<<1$ and
the product of  $\delta^2\beta_1<<1$.

\subsection{Linear regime of long optical pulses (Paraxial approximation)}
It is easy to see that in the case of long pulses (from $ns$ to
$50-60 ps$) $\delta^2<<1$. The values of the dimensionless
dispersion parameter $\beta_1$  range from a small number in the
visible region up to $\beta_1\propto 1-2$ in UV transparency region
of solids, as it was pointed above. This is why we can neglect the
terms multiplied by the small parameter  $\delta^2<<1$ in
(\ref{eqlin13}), (\ref{lin13}), (\ref{VV13}) and (\ref{VVG13}), and
also the terms containing the product of a small parameter with a
parameter of the order of unity $\delta^2(1+\beta_1)<<1$ and
$\delta^2\beta_1$ in (\ref{eqlin13}), (\ref{lin13}). For wave
packets  with a high number of harmonics under the pulse, the
condition $\alpha\delta^2\approx 1$ is usually fulfilled. Thus, for
long pulses or LF, we reduce these equations to the well-known
scalar paraxial approximations:

a. SVEA and VLAE in a laboratory frame:

\begin{eqnarray}
\label{peqlin13}
 -2i\alpha\delta^2\left(\frac{\partial A}{\partial t}+
\frac{\partial A}{\partial z}\right)+ \Delta_{\bot} A =0.
\end{eqnarray}
b. SVEA and VLAE in a Galilean frame:

\begin{eqnarray}
\label{plin13}
 -2i\alpha\delta^2\frac{\partial A}{\partial t'}+
\Delta_{\bot} A=0.
\end{eqnarray}
It is clearly seen that the dynamics of long pulses propagating in the
transparency region of gases and solids is governed by the same
equations that describe the evolution of long pulses in vacuum and are
similar to the optical beam equation. If we turn our attention to the dimension
variables, we will see that there is only one main difference between
propagation of LF in a media with non-stationary optical response and
propagation of LF in vacuum. In paraxial SVEA, the long pulses
propagate with group velocity $v$, while in paraxial VLAE, the
long pulses propagate with velocity of light $c$. Let us present the
initial condition of the paraxial equations (\ref{peqlin13}),
(\ref{plin13}) correspondingly $A^L(x,y,z+t=0)=A_0^L(x,y)$,
$A^G(x,y,t=0)=A_0^G(x,y)$, where $A_0^L$, $A_0^G$ and $A_0^G$ are the
initial amplitude functions in "Laboratory" and "Galilean" coordinates.
Then the solutions of equations (\ref{peqlin13}), (\ref{plin13}) are
well known and they are obtained as a convolution of two inverse
Fourier transforms, the Fourier transform of the initial pulse
(in our case Gaussian pulse) and the spectral presentations of the
Fresnel's kernel in "Laboratory":
\begin{eqnarray}
\label{psol1} A^L(x,y,z+t)=
F^{-1}\left[\hat{A}_0^L(k_x,k_y)\right]\bigotimes
F^{-1}\left[\exp\left(\frac{-i(z+t)}{2i\alpha\delta^2}\left(k_x^2+k_y^2\right)\right)\right]=
\nonumber\\
\\\sqrt{\frac{i\alpha\delta^2}{\pi(z+t)}}
\int\int_{R^2}\left[\exp \left(-\frac{i\alpha\delta^2}{2(z+t)}
\left[(x-x')^2+(y-y')^2 \right]\right)\times
A_0^L(x',y')\right]\,dx'\,dy' ,\nonumber
\end{eqnarray}

"Galilean":
\begin{eqnarray}
\label{psol2} A^L(x,y,t')=
F^{-1}\left[\hat{A}_0^L(k_x,k_y)\right]\bigotimes
F^{-1}\left[\exp\left(-\frac{t'}{2i\alpha\delta^2}\left(k_x^2+k_y^2\right)\right)\right]=
\nonumber\\
\\\sqrt{\frac{i\alpha\delta^2}{\pi t'}}
\int\int_{R^2}\left[\exp \left(-\frac{i\alpha\delta^2}{2t'}
\left[(x-x')^2+(y-y')^2 \right]\right)\times
A_0^L(x',y')\right]\,dx'\,dy'. \nonumber
\end{eqnarray}
As it can be seen from the solutions in the different coordinate
frames, we have exact Fresnel's integrals and phase modulation not
only in time ("Galilean") (\ref{psol2}) but also in space-time
("Laboratory")(\ref{psol1}). These considerations provide an insight
into the real phase modulation of long optical pulses in paraxial
approximation. The pulses obtain equal amounts of space and time
phase modulation under propagation in the media and in vacuum. This
is an important difference from beam propagation, where there is no
group velocity or velocity of light in vacuum in the laboratory
frame and the beam undergoes only spatial modulation. We can
therefore make the following conclusion: The propagation of long
pulses in the region of transparency of a media and in vacuum are
equal with accuracy of the type of velocity of propagation and
undergo Fresnel's diffraction.

\subsection{LB (from few ps to 300-400 fs) and LD (from 200 fs to 30-40 fs) in linear regime }.

We will apply the Fourier method to solve the linear SVEA
(\ref{eqlin13}), (\ref{lin13}) which describe the propagation of
ultrashort optical pulses in media with dispersion and linear VLAE
(\ref{VV13}), (\ref{VVG13}) which govern the propagation of light
pulses in vacuum and dispersionless media. To distinguish between the solutions
of SVEA and VLAE we will mark the Fourier transform of the
amplitude functions of SVEA (\ref{eqlin13}), (\ref{lin13}) in a
Galilean frame with $\vec A_G(k_x,k_y,k_z,t)=F(\vec{A}(x,y,z,t))$,
and in a Laboratory frame with $\vec
A_L(k_x,k_y,k_z,t)=F(\vec{A}(x,y,z,t))$, and the Fourier
transform of the amplitude functions of VLAE (\ref{VV13}),
(\ref{VVG13}) in a Galilean frame with $\vec
B_G(k_x,k_y,k_z,t)=F(\vec{A}(x,y,z,t))$ and in a Laboratory frame with
$\vec B_L(k_x,k_y,k_z,t)=F(\vec{A}(x,y,z,t))$. Applying spatial
Fourier transformation of the components of the amplitude vector
functions $\vec A$, we obtain the following ordinary linear differential
equations in $k_x,k_y,k_z$ space for SVEA:

a. "Laboratory":

\begin{eqnarray}
\label{Fe13}
 -2i\alpha\delta^2\frac{\partial\vec A_L}{\partial t}=
-\left({k_x}^2+{k_y}^2+\delta^2({k_z}^2-2\alpha k_z)\right)\vec
A_L-\delta^2\left(\beta_1+1\right)\frac{\partial^2\vec A_L}{\partial
t^2},
\end{eqnarray}
b. "Galilean":

\begin{eqnarray}
\label{Fl13}
-2i\delta^2\left(\alpha-\left(\beta_1+1\right)k_z\right)\frac{\partial\vec
A_G}{\partial
t}=-\left({k_x}^2+{k_y}^2-\delta^2\beta_1k_z^2\right)\vec
A_G-\delta^2\left(\beta_1+1\right)\frac{\partial^2\vec A_G}{\partial
t^2},
\end{eqnarray}
and the following equations for VLAE:

a. "Laboratory":

\begin{eqnarray}
\label{FV13} -2i\alpha\delta^2\frac{\partial\vec B_L}{\partial t}=-
\left({k_x}^2+{k_y}^2+\delta^2({k_z}^2-2\alpha k_z)\right)\vec
B_L-\delta^2\frac{\partial^2\vec B_L}{\partial t^2}.
\end{eqnarray}
b. "Galilean":

\begin{eqnarray}
\label{FGV13}
 -2i\delta^2\left(\alpha-k_z\right)\frac{\partial\vec B_G}{\partial
 t}=
-\left({k_x}^2+{k_y}^2\right)\vec B_G-\delta^2\frac{\partial^2\vec
B_G}{\partial t^2}.
\end{eqnarray}
We will now denote the square of the sum of the wave vectors with
$\hat{k}^2={k_x}^2+{k_y}^2+\delta^2({k_z}^2-2\alpha k_z)$. We look
for solutions of the kind $\vec
A_L=\vec{A}_L(k_x,k_y,k_z)\exp(i\Omega_L t)$ and $\vec
A_G=\vec{A}_G(k_x,k_y,k_z)\exp(i\Omega_G t)$ for the equations
(\ref{Fe13}), (\ref{Fl13}), and for solutions of the kind $\vec
B_L=\vec{B}_L(k_x,k_y,k_z)\exp(i\Phi_L t)$ and $\vec
B_G=\vec{B}_G(k_x,k_y,k_z)\exp(i\Phi_G t)$ for the equations
(\ref{FV13}) and (\ref{FGV13}) correspondingly. The solutions exist
when $\Omega_L$, $\Omega_G$,  $\Phi_L$ and $\Phi_G$ satisfy the
following quadratic equations:

\begin{eqnarray}
\label{L}
\Omega_L^2+2\frac{\alpha}{\beta_1+1}\Omega_L-\frac{\hat{k}^2}{\delta^2(\beta_1+1)}=0,
\end{eqnarray}
\begin{eqnarray}
\label{G}
\Omega_G^2+2\frac{\left(\alpha-(\beta_1+1)k_z\right)}{\beta_1+1}\Omega_G-
\frac{{k_x}^2+{k_y}^2-\delta^2\beta_1k_z^2}{\delta^2(\beta_1+1)}=0.
\end{eqnarray}

\begin{eqnarray}
\label{VL} \Phi_L^2+2\alpha\Phi_L-\frac{\hat{k}^2}{\delta^2}=0,
\end{eqnarray}
\begin{eqnarray}
\label{VG}
\Phi_G^2+2(\alpha-k_z)\Phi_G-\frac{{k_x}^2+{k_y}^2}{\delta^2}=0.
\end{eqnarray}
The solutions of (\ref{L}),(\ref{G}), (\ref{VL}), and (\ref{VG}) are
similar:

\begin{eqnarray}
\label{OmegaL}
{\Omega_L}^{1,2}=-\frac{\alpha}{\beta_1+1}\pm\sqrt{\frac{\alpha^2}{(\beta_1+1)^2}+\frac{\hat{k}^2}{\delta^2(\beta_1+1)}},
\end{eqnarray}
\begin{eqnarray}
\label{OmegaG}
{\Omega_G}^{1,2}=-\frac{\alpha-(\beta_1+1)k_z}{\beta_1+1}\pm
\sqrt{\frac{\left(\alpha-(\beta_1+1)k_z\right)^2}{(\beta_1+1)^2}+\frac{k_x^2+k_y^2-
\delta^2\beta_1k_z^2}{\delta^2(\beta_1+1)}},
\end{eqnarray}

\begin{eqnarray}
\label{PhiL}
{\Phi_L}^{1,2}=-\alpha\pm\sqrt{\alpha^2+\hat{k}^2/\delta^2},
\end{eqnarray}
\begin{eqnarray}
\label{PhiG}
{\Phi_G}^{1,2}=-(\alpha-k_z)\pm\sqrt{\alpha^2+\hat{k}^2/\delta^2}.
\end{eqnarray}
Now the necessity of a parallel investigation of the propagation of
optical pulses in media with dispersion, in dispersionless media and
in vacuum becomes obvious. Further, we will introduce here the
concept of week and strong dispersion media depending on the value
of dimensionless dispersion parameter $\beta_1$. When $\beta_1<<1$
we have media with week dispersion. It is no difficult to calculate
that in optical transparency region of gases, liquids and solids
materials, this parameter is usually is very small ( $\beta_1<<1$).
For media with week dispersion the solutions of the characteristics
with dispersion (\ref{OmegaL}), (\ref{OmegaG}) are identical with
the solutions for media without dispersion and vacuum (\ref{PhiL}),
(\ref{PhiG}) correspondingly. In media with strong dispersion
$\beta_1$ can reach the values $\beta_1\simeq1-3$ in the UV
transparency region of solids and liquids. In this case, the
solutions for media with dispersion will be slightly modified with
the factor $\beta_1+1$ with respect to the solutions without
dispersion. We consider here the regime of propagation far away from
electronic resonances and the Langmuir frequency in electronic
plasmas, where it is possible to obtain a strongly negative
dispersion parameter $\beta_1\propto-1$. We point here again, that
in the case of LB, when $\beta_1\cong-1$, the amplitude equations
(\ref{svea}) can be transformed into the 3D+1 linear and nonlinear
vector Schrödinger equations \cite{IJMMS}. Generally speaking, the
dispersion parameter $\beta_1$ varies slowly from the visible to the
UV transparency region of the materials from very small values up to
$\beta_1\simeq 1-3$;  this is why it does not influence radically
the solutions and the propagation of optical pulses in linear
regime. The other parameters $\alpha$ and $\delta^2$ change
significantly. For example $\alpha$ varies from $10^1$ to $10^3$
while $\delta^2$ varies from $10^{-2}-10^{-4}$ for LF to $10^0$ for
LB and $10^2-10^4$ for LD. This is why in the next paragraph we will
investigate more precisely the solutions of the equations for media
with week dispersion, dispersionless media and vacuum (\ref{PhiL}),
(\ref{PhiG}), as we expect that the solutions for media with strong
dispersion (UV transparency region of solids and liquids)
(\ref{OmegaL}), (\ref{OmegaG}) will be only slightly modified by the
factor $\beta_1+1$. As we obtained the solutions for the
characteristics (\ref{OmegaL}), (\ref{OmegaG}), (\ref{PhiL}) and
(\ref{PhiG}), the solutions of the corresponding linear differential
equations  SVEA (\ref{Fe13}), (\ref{Fl13}) and VLAE (\ref{FV13}),
(\ref{FGV13}) in the $k$-space become:

a. Solution of SVEA in the k-space and a laboratory coordinate frame:

\begin{eqnarray}
\label{SOL} \vec{A_L}=\vec{A_L}(k_x,k_y,k_z,t=0)\times\nonumber\\
\exp\left(i\left(-\frac{\alpha}{\beta_1+1}
\pm\sqrt{\frac{\alpha^2}{(\beta_1+1)^2}+\frac{\hat{k}^2}{\delta^2(\beta_1+1)}}\right)t\right).
\end{eqnarray}
b. Solution of SVEA in the k-space and a Galilean coordinate frame:

\begin{eqnarray}
\label{SOG}\vec{A_G}=\vec{A_G}(k_x,k_y,k_z,t=0)\times\nonumber\\
\exp\left(-i\left(-\frac{\alpha-(\beta_1+1)k_z}{\beta_1+1}
\pm\sqrt{\frac{\left(\alpha-(\beta_1+1)k_z\right)^2}{(\beta_1+1)^2}+\frac{k_x^2+k_y^2-
\delta^2\beta_1k_z^2}{\delta^2(\beta_1+1)}}\right)t\right).
\end{eqnarray}
c. Solution of VLAE in k-space and laboratory coordinate frame:

\begin{eqnarray}
\label{BSOL}
\vec{B_L}=\vec{B_L}(k_x,k_y,k_z,t=0)\exp\left(i\left(-\alpha\pm\sqrt{\alpha^2+\hat{k}^2/\delta^2}\right)t\right).
\end{eqnarray}
d. Solution of VLAE in the k-space and a Galilean coordinate frame:

\begin{eqnarray}
\label{BSOG}\vec{B_G}=\vec{B_G}(k_x,k_y,k_z,t=0)\exp
\left(-i\left((\alpha-k_z)\pm\sqrt{\alpha^2+\hat{k}^2/\delta^2}\right)t\right).
\end{eqnarray}
It is natural that the solutions (\ref{BSOL}) and (\ref{BSOG}) of
equations (\ref{FV13}) and (\ref{FGV13}) should identical be equal
with accuracy to a wave number in the z direction. This follows from
the Fourier transform of such evolution equations and leads to only
one difference between the solutions in the real space - the motion
of the pulse in the z direction in a "Laboratory" frame and its
stationarity in a "Galilean" frame. We deal here only with initial
functions for the amplitude envelopes of the electric field that are
localized in space and time. Thus, the images of these functions
after Fourier transform in the $k_x,k_y,k_z$ space also are
localized functions. The solutions of our amplitude equations in $k$
space (\ref{SOL}), (\ref{SOG}),(\ref{BSOL}) and (\ref{BSOG}) are the
product of the initial localized in $k_x,k_y,k_z$-space functions
and the new spectral kernels, which are periodic  (different from
Fresnel's one). The product of a localized function and a periodic
function is also a function localized in the $k_x,k_y,k_z$ space.
Therefore, the solutions of our amplitude equations in $k$ space
(\ref{BSOL}), (\ref{BSOG}),(\ref{SOL}) and (\ref{SOG}) also are
localized functions in this space and we can apply the inverse
Fourier transform to obtain again localized solutions in the
$x,y,z,t$ space. More precisely, we use the convolution theorem to
present our solution in the real space as a convolution of inverse
Fourier transform of the initial pulse with the inverse Fourier
transforms of the new spectral kernels:

a. Solution of SVEA (\ref{eqlin13}) in a laboratory coordinate frame:

\begin{eqnarray}
\label{L1} \vec{A}(x,y,z,t)=F^{-1}\left(\vec{A_L}(k_x,k_y,k_z,t=0)\right)\otimes\nonumber\\
F^{-1}\left(\exp\left(i\left(-\frac{\alpha}{\beta_1+1}
\pm\sqrt{\frac{\alpha^2}{(\beta_1+1)^2}+\frac{\hat{k}^2}{\delta^2(\beta_1+1)}}\right)t\right)\right).
\end{eqnarray}
b. Solution of SVEA (\ref{lin13}) in a Galilean coordinate frame:

\begin{eqnarray}
\label{LG}\vec{A}(x,y,z',t')=F^{-1}\left(\vec{A_G}(k_x,k_y,k_z,t=0)\right)\otimes\nonumber\\
F^{-1}\left(\exp\left(-i\left(-\frac{\alpha-(\beta_1+1)k_z}{\beta_1+1}
\pm\sqrt{\frac{\left(\alpha-(\beta_1+1)k_z\right)^2}{(\beta_1+1)^2}+\frac{k_x^2+k_y^2-
\delta^2\beta_1k_z^2}{\delta^2(\beta_1+1)}}\right)t\right)\right).
\end{eqnarray}
a. Solution of VLAE (\ref{VV13}) in the laboratory coordinate frame:

\begin{eqnarray}
\label{L2}
\vec{A}(x,y,z,t)=F^{-1}\left(\vec{B_L}(k_x,k_y,k_z,t=0)\right)\otimes\nonumber\\
F^{-1}\left(\exp\left(i\left(-\alpha\pm\sqrt{\alpha^2+\hat{k}^2/\delta^2}\right)t\right)\right).
\end{eqnarray}
b. Solution of VLAE (\ref{VVG13})in a Galilean coordinate frame:

\begin{eqnarray}
\label{LG2}
\vec{A}(x,y,z',t')=F^{-1}\left(\vec{B_G}(k_x,k_y,k_z,t=0)\right)\otimes\nonumber\\
F^{-1}\left(\exp
\left(-i\left((\alpha-k_z)\pm\sqrt{\alpha^2+\hat{k}^2/\delta^2}\right)t\right)\right),
\end{eqnarray}
where with $F^{-1}$ we denote the spatial three-dimensional inverse
Fourier transform and with $ \otimes$ we denote the convolution
symbol. The difference between the Fresnel's integrals (\ref{psol1})
and (\ref{psol2}), describing propagation of optical beams in linear
regime of propagation, and the new integrals (\ref{L1}), (\ref{LG}),
(\ref{L2}) and (\ref{LG2}), solutions of the linear evolution
equations (\ref{eqlin13}), (\ref{lin13}), (\ref{VV13}) and
(\ref{VVG13}) is quite obvious. In addition in the new spectral
kernels there are three dimensionless parameters, $\alpha$, which
determines the number of harmonics under the pulse with accuracy
$2\pi$, $\delta^2$, which gives the relation between the initial
longitudinal and transverse dimensions of the optical pulse, and
$\beta_1$, which accounts for the influence of the second-order
dispersion. We will focus mainly on the SVEA, and will fixed
$\alpha$ to be always a large number $\alpha>>1$. As it was pointed
out above, the condition $\alpha>>1$ is not necessary for VLAE so
that we can investigate pulses with longitudinal duration of order
of the carrier wavelength in vacuum and in dispersionless media. To
analyze more precisely the influence of the other two parameters,
$\delta^2$ and $\beta_1$ on the evolution of the initial pulse we
will rewrite the expression for the spectral kernel (\ref{OmegaL})
of the solutions (\ref{L1}) of equation (\ref{eqlin13}) in the
following form:

\begin{eqnarray}
\label{OmL}
{\Omega_L}^{1,2}=-\frac{\alpha}{\beta_1+1}\nonumber\\
\pm\sqrt{\frac{\alpha^2}{(\beta_1+1)^2}+\frac{1}{\delta^2(\beta_1+1)}(k_x^2+k_y^2)+
\frac{1}{\beta_1+1}k_z^2-\frac{2\alpha}{\beta_1+1}k_z}.
\end{eqnarray}
As $\alpha$ is a constant, the diffraction widening will be
determined by the second and third term under the square root in
(\ref{OmL}):
\begin{eqnarray}
\label{kern} \frac{1}{\delta^2(\beta_1+1)}(k_x^2+k_y^2)+
\frac{1}{\beta_1+1}k_z^2,
\end{eqnarray}
which determines the transverse and longitudinal diffraction
and dispersion widening of the pulses. We pointed out above that the
dispersion parameter varies very slowly within the limits
$0\leq\beta_1<10^1$, while the relations between the transverse and
longitudinal part varies significantly $10^{-4}<\delta^2<-10^4$. This is why
we will estimate mainly the influence of the different
values of $\delta^2$ and $\alpha^2$ on the diffraction widening. We
will consider investigate the following basic cases:

a/Long pulses, when $\delta^2<<1$. It is easy to estimate from
(\ref{kern}), that the transverse enlargement ${k_x}^2+{k_y}^2$ will
dominate significantly over the longitudinal one $k_z^2$ as:
\begin{eqnarray}
\label{KL} \frac{1}{\delta^2(\beta_1+1)}>>1 ;\
\frac{1}{\beta_1+1}\cong1/2.
\end{eqnarray}
A more precise mathematical analysis will give us the standard
Fresnel's  spectral kernel and paraxial approximation when
$\beta_1<<1$. In the case of long pulses we have also
$\alpha\delta^2\sim 1$ and it is seen from equations
(\ref{Fe13}) and (\ref{Fl13}) that we can obtain a similar
diffraction length $(z_{diff}=k_0r^{2}_{\perp})$, as in the case of an
optical beam. The difference is only with the factor $\beta_1+1$, while
for the cases of media with week dispersion, dispersionless media and
vacuum, there is exact coincidence between the diffraction of long
pulses and diffraction of optical beams.

b/LB: $\delta^2\simeq 1$. In the case of optical pulses with
relatively equal transverse and longitudinal size, we obtain equal
coefficients in front of the transverse ${k_x}^2+{k_y}^2$ and
longitudinal $k_z^2$ diffraction terms:
\begin{eqnarray}
\label{KB} \frac{1}{\delta^2(\beta_1+1)}\cong 1/2 ;\
\frac{1}{\beta_1+1}k_z^2\cong1/2.
\end{eqnarray}
Hence, the diffraction and dispersion transverse enlargement will be
reduced with the factor $\delta^2(\beta_1+1)$ with respect to the
diffraction of long pulses and Fresnel's diffraction. The LB
widening will be further reduced if the square of the number of
harmonics multiplied by $4\pi^2$ - $\alpha^2$ exceeds the square of
the sum of the normalized wave vectors $(1/2)(k_x^2+k_y^2+k_z^2) $.
For example, the main part in the Fourier integral of one normalized
localized Gaussian pulse gives values of the wave vectors in the
range $-4<k_{x,y,z}<4$ and the  sum is of the order of
$(1/2)(k_x^2+k_y^2+k_z^2)\sim 20-30$. For a pulse with many
harmonics under the envelope, $\alpha^2$ can reach
$\alpha^2\sim10^{2-4}$. Thus, the square of the number of harmonics
$\alpha^2$ under the pulse starts to prevail over the diffractive
terms $\hat{k}^2$ in the Fourier integrals (\ref{SOL}) and
(\ref{SOG}). Therefore, the two conditions, $\delta^2\simeq1$ and
$\alpha^2>>1$, open up a unique possibility for a considerable
reduction of the diffractive effects in the dynamics of LB and for a
relatively stable linear propagation of LB as compared with long
pulses and light beams.

In addition, we will point here to an important asymptotical behavior
of LB: When $\alpha^2$ is small (pulses with only one two
harmonics under the envelope) and $\beta_1<<1$ (media with week
dispersion, dispersionless media and vacuum), the spectral kernels
of the new equations tend to the asymptotical value $\sim
\exp\left(i(\sqrt{k_x^2+k_y^2+(k_z+\alpha)^2})t\right)\cong\exp\left(i(|k|t)\right)$,
which is actually the spectral kernel of the 3D wave equation. This is why for optical pulses with
only one or two harmonics under the
envelope we can expect diffraction similar to the typical
diffraction of the 3D wave equation, with the appearance of observing  of internal and
external fronts and a significant widening of the pulse.

c/ Light disks: This is the case when the longitudinal size $z_0$
is mush shorter than the transverse size $r_\bot$ and $\delta^2>>1$.
As it was indicated above, the typical time region for such pulses is
$30-40 fs<t_0<200-300 fs$. We determine the lower limits of this
relation from the conditions $\alpha^2>>1$, i.e a large number of harmonics under
the envelope. This condition still holds true for pulses in the visible and UV regions with time duration $30-40 fs$.
The dimensionless parameter
in front of the transverse diffraction and dispersion
$\left({k_x}^2+{k_y}^2\right)$ will be of the order of:
\begin{eqnarray}
\label{KLL} \frac{1}{\delta^2(\beta_1+1)}<<1,
\end{eqnarray}
while for the coefficient in front of the longitudinal diffraction
$k_z^2$ we have:

\begin{eqnarray}
\label{KN} \frac{1}{(\beta_1+1)}\cong 1/2 .
\end{eqnarray}
We thus see that the transverse enlargement is of the order of
$\delta^2(\beta_1+1)$, or negligible as compared with LB, and smaller by a factor of about
$(\delta^2(\beta_1+1))^2$ than in the cases of long pulses and
Fresnel's diffraction. The longitudinal enlargement $k_z^2$ is of the
order of that for LB. To summarize the results of this section, we can we expect that the
transverse diffraction and dispersion enlargement of
LB should be smaller by a factor $\delta^2(\beta_1+1)$ than those of LF,
while the transverse diffraction and dispersion enlargement of LD should be smaller by a factor of
$(\delta^2(\beta_1+1))^2$ than those of long
pulses and paraxial approximation. Practically no
transverse enlargement of LD would be observed over long
distances, namely, more than tens or hundreds of kilometers.

\section{Dynamics of LF, LB and LD in media with weak dispersion,
dispersionless media and vacuum}

In the previous section we first found the solutions of SVEA
(\ref{VV13}) and VLAE (\ref{VVG13}) in the $k$ space
(\ref{SOL}),(\ref{SOG}),(\ref{BSOL}) and (\ref{BSOG}). The dynamics
of the solutions under different initial conditions in the $(x,y,z,t)$,
space can be found by applying an inverse Fourier transform to
the solutions in the $k$ space (\ref{L1}), (\ref{L2}), (\ref{LG}) and
(\ref{LG2}). As it was pointed out before, the dispersion parameter varies very
slowly in the limits $0\leq\beta_1<10^1$, while the relations
$\delta^2$ and $\alpha$  vary significantly. We will, therefore,
study the VLAE, as the solutions of SVEA for media with strong
dispersion(\ref{L1}) and (\ref{L2}) differ from the solutions of
VLAE (\ref{LG}) and (\ref{LG2}) only by the factor $\beta_1+1$. We should remind here
that we investigate in this paper only the case of the
region of transparency of the materials with positive dispersion. In
the beginning we will fix the dimensionless parameter $\alpha$
(number of harmonics multiplied by $2\pi$) to be a large number number
$\alpha=40$ and consider the following basic cases: dynamics of LF
($\delta^2<<1$), light bullets ($\delta^2=1$ and light disks
$\delta^2>>1$. We compare the evolutions of LF, LB and LD with the
solution of a Gaussian beam in the framework of the standard paraxial
equation of linear optics, written in a laboratory coordinate frame
(\ref{peqlin13}) with initial condition of the type:

\begin{eqnarray}
\label{PGAUS}
\vec{A}=A_x\vec{x};\nonumber\\
A_x(x,y,z=t=0)=\exp\left(-\frac{x^2+y^2}{2}\right).
\end{eqnarray}
The evolutions of the initial condition (\ref{PGAUS}) governed by
the paraxial equation (\ref{peqlin13}) is described by the
Fresnel's integral  or can be found by numerical calculation of the inverse
Fourier transform of the solution in the ($k_x,k_y$)-space
(\ref{psol1}). The solutions of the paraxial equation (\ref{peqlin13})
with initial condition (\ref{PGAUS}) on a normalized
distance $4\pi/15$ is illustrated in Fig.1.

\begin{figure}[t]\label{Fig1}
\begin{center}
\includegraphics[width=120mm,height=50mm]{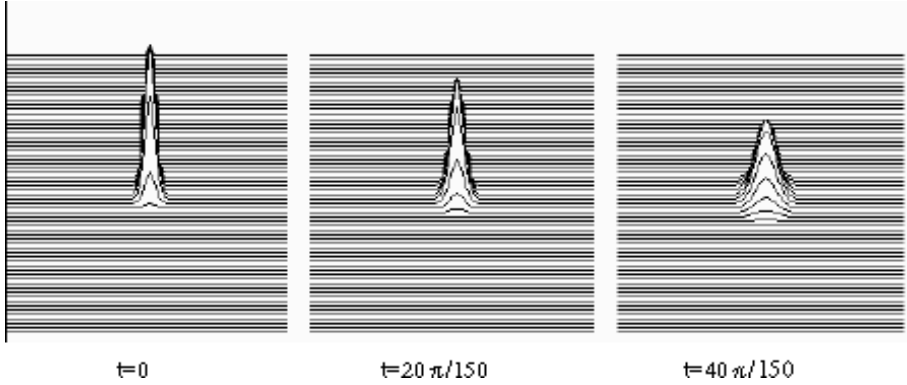}
\caption{Evolution of a Gaussian beam with initial condition
$A_x(x,y,z=t=0)=\exp\left(-\frac{x^2+y^2}{2}\right)$ governed by
the $2D$ paraxial equation (\ref{peqlin13}). The transverse size
(the spot) grows approximately twice over the normalized distance
$z=4\pi/15$.}
\end{center}
\end{figure}

\subsection{Evolution of long optical pulses (light filaments)}

The dynamics of optical pulses in media with week dispersion,
dispersionless media and vacuum is governed by the VLAE
(\ref{VV13}). In the case of a long pulse we can express the initial
conditions as follows:

\begin{eqnarray}
\label{PLONG}
\vec{A}=A_x\vec{x};\ \alpha=40;\  \delta^2=\frac{r_{\bot^2}}{z_0^2}=\frac{1}{81}, \nonumber\\
A_x(x,y,z,t=0)=\exp\left(-\frac{x^2+y^2+\delta^2z^2}{2}\right).
\end{eqnarray}
The solution of VLAE in a laboratory coordinate frame (\ref{L2}) for an
optical pulse whose longitudinal size is nine times as large as the
transverse one (\ref{PLONG}) is calculated by using the numerical FFT
technique. The result is presented in Fig.2. The propagation distance
is the same as in the previous case of a light beam $4\pi/15$.
Figure 3. illustrates the spot size ($x,y$ plane) of the
pulse. The analysis of the solution reveals the same evolution
dynamics as in the case of an optical beam, with the diffraction length being
approximately equal to the diffraction length of an optical beam $4\pi/15$.
A more detailed investigation discloses the following natural dependence
on the dimensionless parameters $\alpha$ and $\delta^2$ in the case of
a long pulse: as the number of harmonics under the
pulse $\alpha$ increases, together with the longitudinal size with respect to the transverse
one, the VLAE solutions (\ref{L2}) approach the paraxial equation solutions (\ref{peqlin13}).

\begin{figure}[h]\label{Fig2}
\begin{center}
\includegraphics[width=120mm,height=50mm]{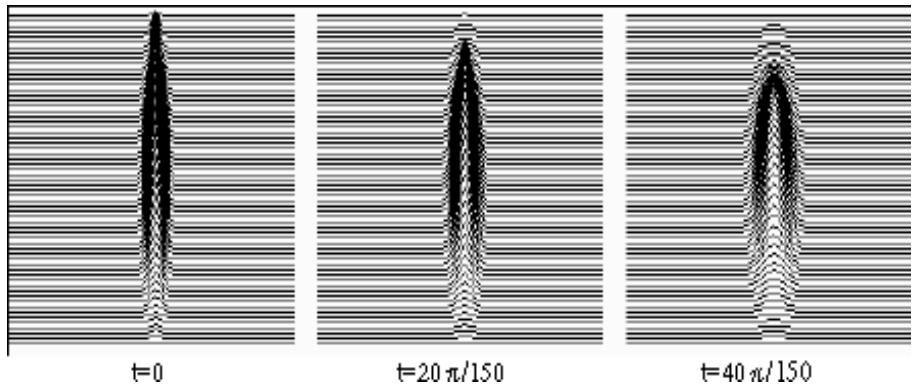}
\caption{Evolution of a long Gaussian  pulse with initial condition
$A_x(x,y,z,t=0)=\exp\left(-\frac{x^2+y^2+\delta^2z^2}{2}\right)$,
$\alpha=40;\delta=1/9$ governing by the VLAE (\ref{VV13}). The $xz$ plane of the pulse is
presented.}
\end{center}
\end{figure}

\begin{figure}[h]\label{Fig3}
\begin{center}
\includegraphics[width=120mm,height=50mm]{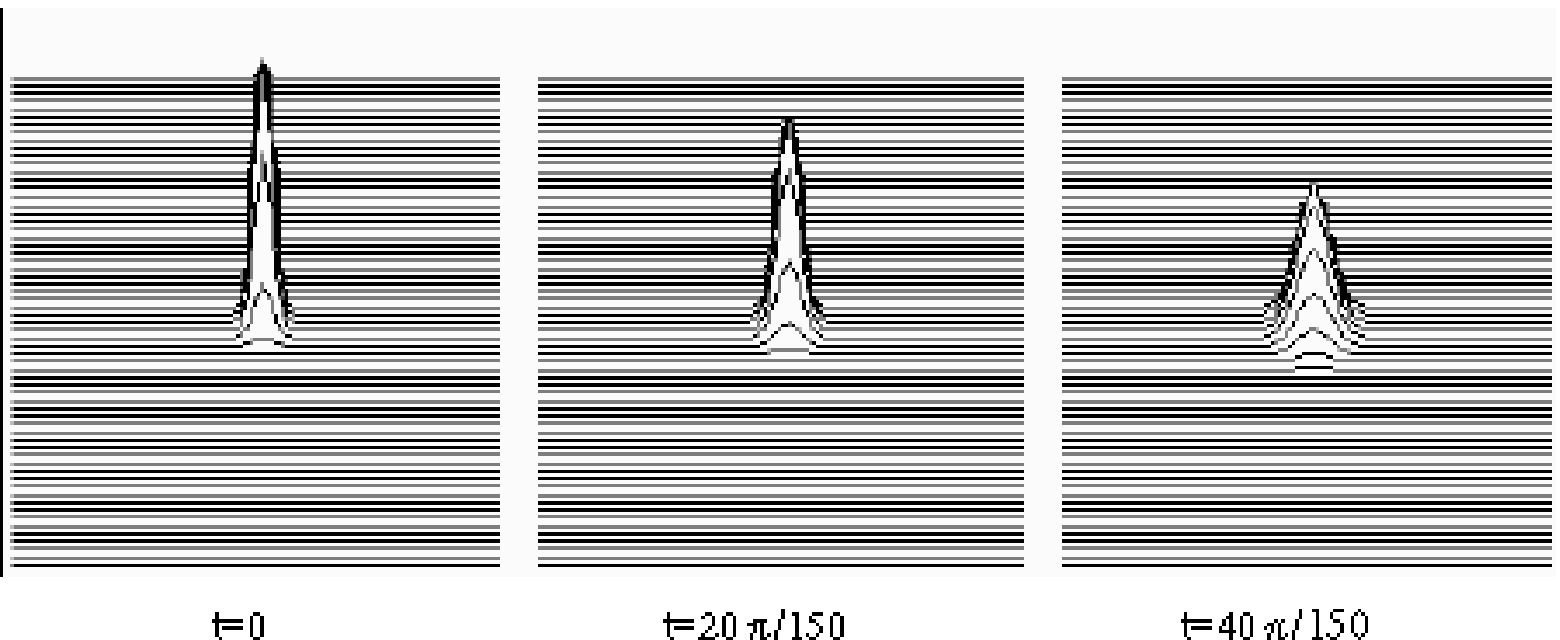}
\caption{View of the spot, $x,y$ plane, of the same long Gaussian
pulse as in Fig.2. The transverse size grows approximately twice
over the same normalized distance $z=4\pi/15$ as in the
case of an optical beam.}
\end{center}
\end{figure}

\subsection{Propagation of light bullets in linear regime}
The evolution of LB in media with week dispersion, dispersionless
media and vacuum is governed by the same VLAE (\ref{VV13}). The
shape of the LB is symmetric in the $x$, $y$ and $z$ plane, so that
the initial Gaussian profile can be written as:

\begin{eqnarray}
\label{LBul}
\vec{A}=A_x\vec{x};\ \alpha=40;\  \delta^2=\frac{r_{\bot^2}}{z_0^2}=1, \nonumber\\
A_x(x,y,z,t=0)=\exp\left(-\frac{x^2+y^2+z^2}{2}\right).
\end{eqnarray}

From the qualitative analysis presented in the previous section, we
can expect a significant reduction of the widening of LB (of the
order of $1/\delta^2$) with respect to long pulses or optical beams.
The solution of VLAE (\ref{L2}) for LB on a distance $5\pi$,
calculated again by means of the FFT technique, is illustrated in
Fig.4. in the $x,z$ plane. One can only see a negligible enlargement
of the pulse along a considerable distance. Figure 4. presents the
solution (\ref{L2}) of the VLAE (\ref{VV13}), with initial
conditions of the type (\ref{LBul}). If we consider a large distance
and times $z=t>5\pi$, the LB will go out äî the grid. This is the
reason why the next calculations of LB and LD evolution over long
distances are performed in framework of the solutions (\ref{LG2}) of
the VLAE (\ref{VVG13}) written in a Galilean coordinate system under
similar initial conditions:

\begin{eqnarray}
\label{LBG}
\vec{A}=A_x\vec{x};\ \alpha=40;\  \delta^2=\frac{r_{\bot^2}}{z_0^2}=1, \nonumber\\
A_x(x,y,z',t'=0)=\exp\left(-\frac{x^2+y^2+z'^2}{2}\right).
\end{eqnarray}
The LB evolution in a Galilean frame over a long normalized distance
$t=15\pi$, where the transverse width of the pulse is approximately
doubled, is illustrated in Figure 5. The analytical solutions
confirm the results of the qualitative analysis: the LB diffraction
enlargement is significantly smaller than the enlargement of long
optical pulses and optical beams. Let us remark once again that this
result is only correct if the number of harmonics under the pulse,
multiplied by $2\pi$, (dimensionless parameter $\alpha$) is large.

\begin{figure}[h]\label{Fig4}
\begin{center}
\includegraphics[width=80mm,height=50mm]{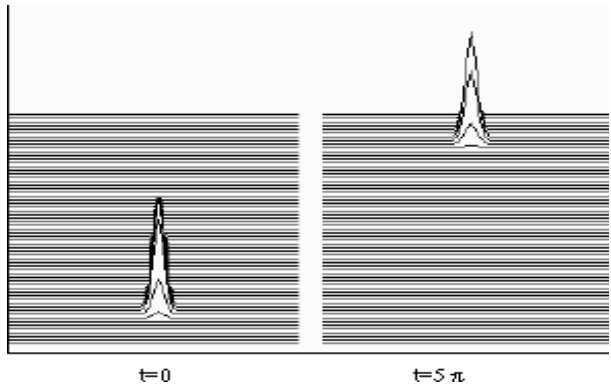}
\caption{Evolution of a Gaussian LB under initial condition
$A_x(x,y,z,t=0)=\exp\left(-\frac{x^2+y^2+z^2}{2}\right)$,
$\alpha=40;\delta=1$ governed by the VLAE (\ref{VV13}). We present
the $xz$ plane of the pulse. A significantly smaller diffraction
enlargement is observed, as compared with the cases of LF and light beams.}
\end{center}
\end{figure}
\begin{figure}[h]\label{Fig5}
\begin{center}
\includegraphics[width=120mm,height=110mm]{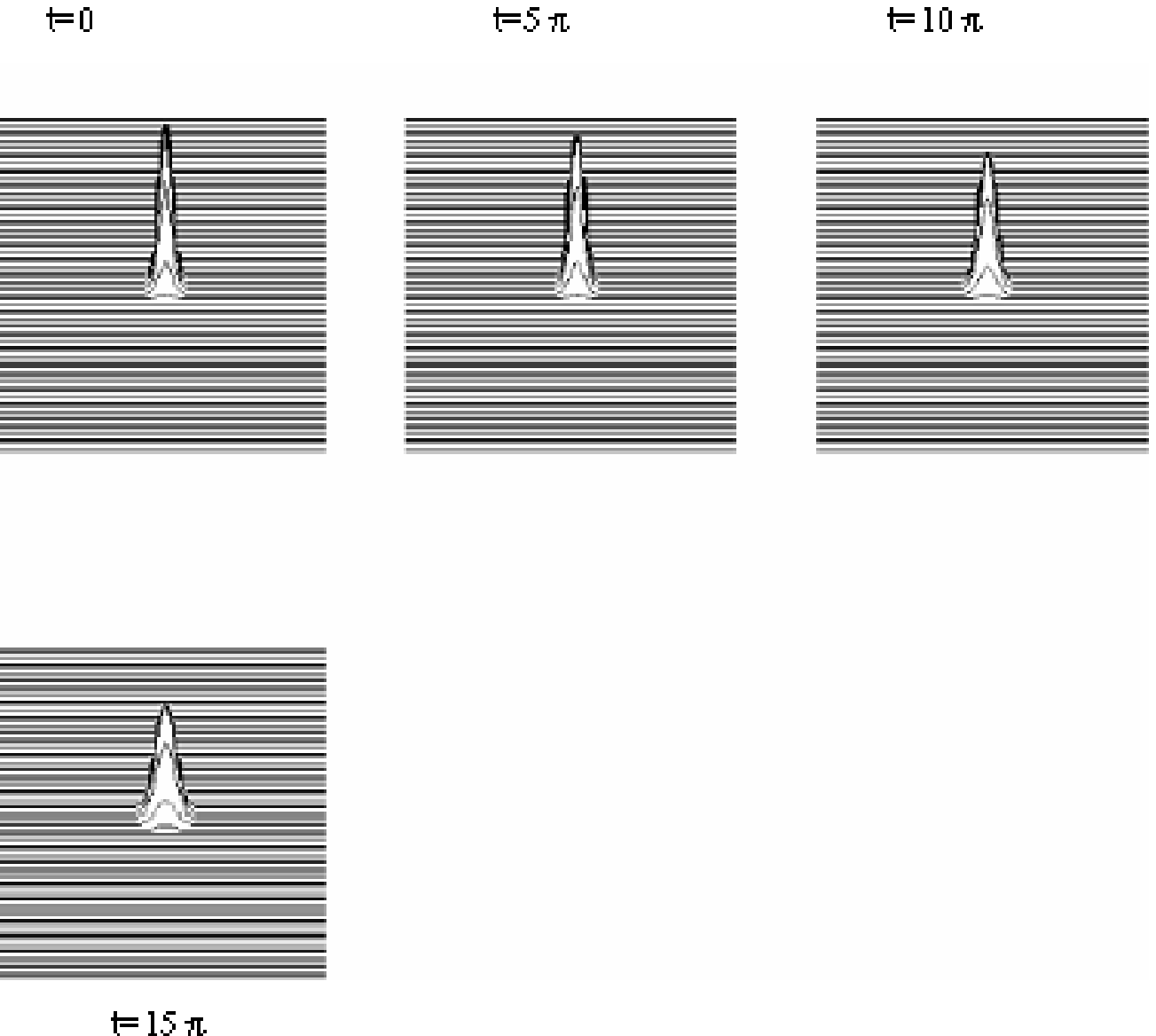}
\caption{Evolution of the same Gaussian LB as in Fig.4 , governed
now by the VLAE in a Galilean coordinate frame (\ref{VVG13}). The
$xy$ plane of the pulse is presented. The LB doubles its transverse
(and longitudinal) size over the normalized distance-time
$t'=z'=15\pi$. This is approximately $10^2$ larger than the
diffraction length of LF and light beams.}
\end{center}
\end{figure}

\subsection{Dynamics of of light disks in linear regime. Difractionless pulses}
As we mentioned in the beginning, producing optical pulses with small longitudinal and
large transverse size, while at the same time keeping a large number of harmonics under
the pulse, no longer presents significant experimental difficulties. This can easily be
realized in the optical region for pulses with time duration from 200-300 fs up to 50-60 fs.
We will again consider the propagation of LD in the framework of the solutions
(\ref{LG2}) of the VLAE (\ref{VVG13}) in a Galilean coordinates under
initial conditions in the form:

\begin{eqnarray}
\label{LLBG}
\vec{A}=A_x\vec{x};\ \alpha=40;\  \delta^2=\frac{r_{\bot^2}}{z_0^2}=81, \nonumber\\
A_x(x,y,z',t'=0)=\exp\left(-\frac{x^2+y^2+z'^2}{2}\right).
\end{eqnarray}
The results of the calculations of the solutions (\ref{LG2}) using
FFT and inverse FFT are presented in Figure 6. Again, as one should expect bearing in mind
the analysis in the previous section, the diffraction widening is practically absent.
The evolution of the optical disk presented in Figure 6. demonstrates that its shape is preserved over the
normalized distance and for time $t=375\pi$. This is larger by four orders of magnitude, than in the cases
of diffraction of light beams and long optical pulses.
\begin{figure}[h]\label{Fig6}
\begin{center}
\includegraphics[width=120mm,height=110mm]{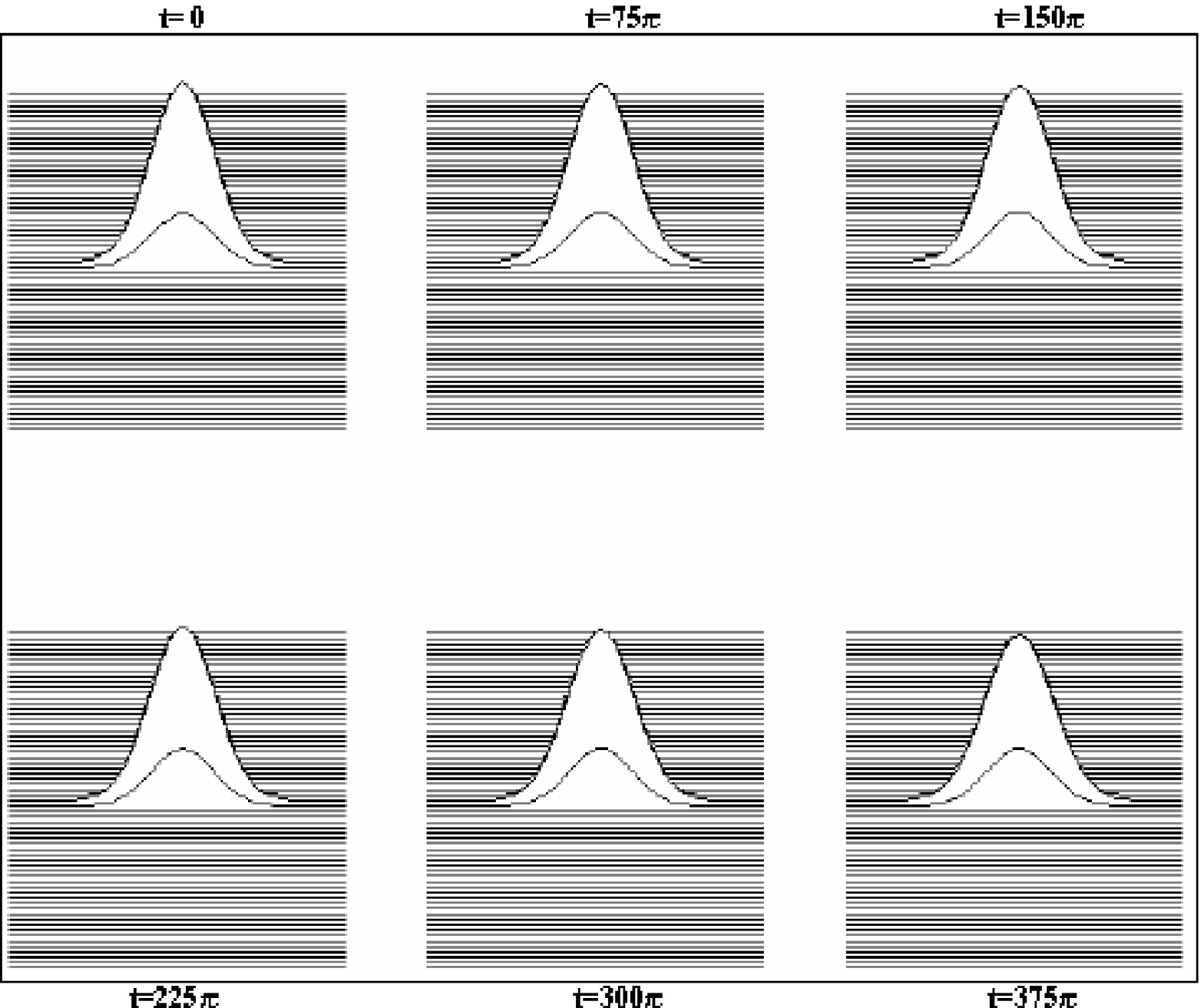}
\caption{Evolution of an optical disk governed now by the VLAE in a
Galilean coordinate frame (\ref{VVG13}). The $xy$ plane of the pulse
presented. The LD preserves its shape over a distance-time $375\pi$,
which is larger by $10^4$ than the standard diffraction length
$z_{diff}=k_0r_{\bot}^2$ of light beams and long optical pulses. We
call this type of optical pulses diffractionless.}
\end{center}
\end{figure}
This is why we refer to optical disks as being diffractionless
pulses. The numerical experiments demonstrate that such pulses can
propagate in media with week dispersion, in dispersionless media and
in vacuum while conserving their shape over a distance of more than
hundred km.

\section{Conclusion}

The method applied of amplitude envelopes give us the possibility to
investigate and compare the propagation of optical pulses in media
with strong dispersion with this in media with week dispersion, in
dispersionless media and in vacuum. In the case of media with
dispersion, we obtained an integro - differential nonlinear
equation, describing propagation of optical pulses whose time
duration is of the order of optical period of the carrier frequency,
and also of pulses with many harmonics under the pulse. In the case
of slowly varying amplitudes (many harmonics under the pulse) we
reduced this amplitude integro - differential equation to amplitude
vector nonlinear differential equations and obtained different
orders of dispersion of the linear and nonlinear susceptibility. In
the second case, propagation of optical pulses in dispersionless
media and vacuum, we obtained an amplitude equation witch is valid
in both cases, namely, pulses with many harmonics and pulses with
only one-two harmonics under the envelope.  We normalized these
amplitude equations and obtained five dimensionless parameters
determining different linear and nonlinear regimes of propagation of
the optical localized waves. For optical pulses with small
transverse and large longitudinal size (optical filaments) we
obtained the well-known paraxial approximation, while for the case
of optical pulses with a relatively equal transverse and
longitudinal size (the so-called light bullets) and for the case of
large transverse and small longitudinal size (light disks), we
obtained new non-paraxial nonlinear amplitude equations. When the
optical field is with low intensity, we reduced the nonlinear
amplitude vector equation governing the evolution of LB and LD to
linear amplitude equations. Surprisingly, in linear regime the
normalized amplitude equations for media with dispersion and the
amplitude equations in dispersionless media and vacuum are identical
with accuracy of material constants. The linear equations were
solved using the of Fourier transform technique. One unexpected new
result is the relative stability of LB and LD and the significant
reduction of the LB and LD diffraction enlargement with respect to
the case of long pulses in linear regime of propagation. It is
important to emphasize particularly the case of LD which are
practically diffractionless over long distances exceeding hundred
kilometers.

\section{Acknowledgements}

This work is partially supported by Bulgarian Science Foundation
under grant F 1515/2005.

\end{document}